\documentclass[twocolumn,prl,float]{revtex4}  

\usepackage{graphicx}
\usepackage{float}

\usepackage{epsfig}

\begin{document}

\title{ Self-assembly of orthorhombic $Fddd$ network in  simple
  one-component liquids} \author{Lorenzo Agosta$^{1}$, Alfredo
  Metere$^{2*}$, Peter Oleynikov $^{3}$ and Mikhail Dzugutov$^4$}

\affiliation{$^1$ Department of Materials and Environmental Chemistry,
  Stockholm University, S-10691 Stockholm, Sweden\\$^2$ International
  Computer Science Institute, 1947 Center St., Berkeley, CA
  94704\\$^3$ Shanghai Tech University, School of Physical Science and
  Technology, Shanghai, China \\ $^4$ Dept. of Chemistry,
  $\AA$ngstr\"{o}m laboratoriet, Uppsala Universitet, 751 21 Uppsala,
  Sweden}

\begin{abstract}

  Triply periodic continuous morphologies arising a result of the
  microphase separation in block copolymer melts have so far never
  been observed self-assembled in systems of particles with
  spherically symmetric interaction.  We report a molecular dynamics
  simulation of two simple one-component liquids which self-assemble
  upon cooling into equilibrium orthorhombic continuous network
  morphologies with the $Fddd$ space group symmetry reproducing the
  structure of those observed in block copolymers. The finding that
  the geometry of constituent molecules isn't relevant for the
  formation of triply periodic networks indicates the generic nature
  of this class of phase transition.

\end{abstract}

\date{\today}


\maketitle

The concept of microphase separation in fluids was introduced more
than fifty years ago by Lebowitz and Penrose \cite{leb}. They
conjectured that amending the pair potential in the van der Waals
model with an additional long-range repulsion can cause the
macroscopic separation of liquid and gas phases in the spinodal domain
to break into mesoscopic-scale patterns. In a separate development, it
was suggested \cite{scri} that fluids of immiscible macromolecules may
form mesoscopic-scale structures composed of  continuous
percolating domains separated by minimal surfaces. Moreover, these
microphases were conjectured to form morphologies with long-range
translational order.  These microphase transitions have been observed
in block co-polymer melts producing a variety of multiply continuous
periodic morphologies which are both elegant and remarkably useful
\cite{meuler}. As a surprising discovery, the orthorombic $Fddd$
network was observed \cite{bail, epps}, the first non-cubic
structure produced in soft-matter systems, raising the discussion
about the origin of that symmetry breaking.

The self-assembly of the triply periodic networks in block copolymer
melts is thought to be controlled by the minimisation of the
inter-domain surface area and the conformation entropy of the
molecular chains. Whether such a network can self-assemble in a system
of particles with spherically symmetric interaction as a result of the
liquid-gas microphase separation remains a question of both conceptual
and practical interest. Landau theory and density-functional theory
\cite{ciach,edelman} predicted that triply periodic networks can be
stabilised by the pair potentials with short-range attraction and
long-range repulsion (SALR).  The interparticle potentials in
colloidal systems of spherical particles can be tuned to approximate
those conjectured from the theoretical models \cite{isr}.  However,
despite the extensive experimental efforts, triply periodic
morphologies have never been observed in colloidal systems.
\cite{klix}.

Simulations using particles are indispensable for testing the ability
of the theoretically conjectured SALR pair potentials to produce
self-assembly of triply periodic networks.  These networks, however,
have so far never been observed self-assembling in simulations using
the SALR pair potentials \cite{camp,sciortino,char}.  Moreover, it was
concluded, based on the density-functional calculations
\cite{edelman}, that if one starts with a random particle
configuration, such a system, due to the complexity of its free-energy
landscape, wouldn't be able to reach the minimum representing a triply
periodic network. Thus, the conceptually significant question of
whether a simple system of identical particles can self-assemble into
a triply periodic network remains open.

This Letter reports a molecular-dynamics simulation of two different
systems of identical particles with spherically-symmetric interaction
which demonstrate microphase separation transitions upon cooling from
a uniform liquid state. The phase diagrams for both systems are found
to include domains of density and temperature where they self-assemble
into equilibrium triply periodic morphologies possessing orthorhombic
$Fddd$ space-group symmetry.

\begin{figure} [h!] 
\includegraphics[width=6.cm]{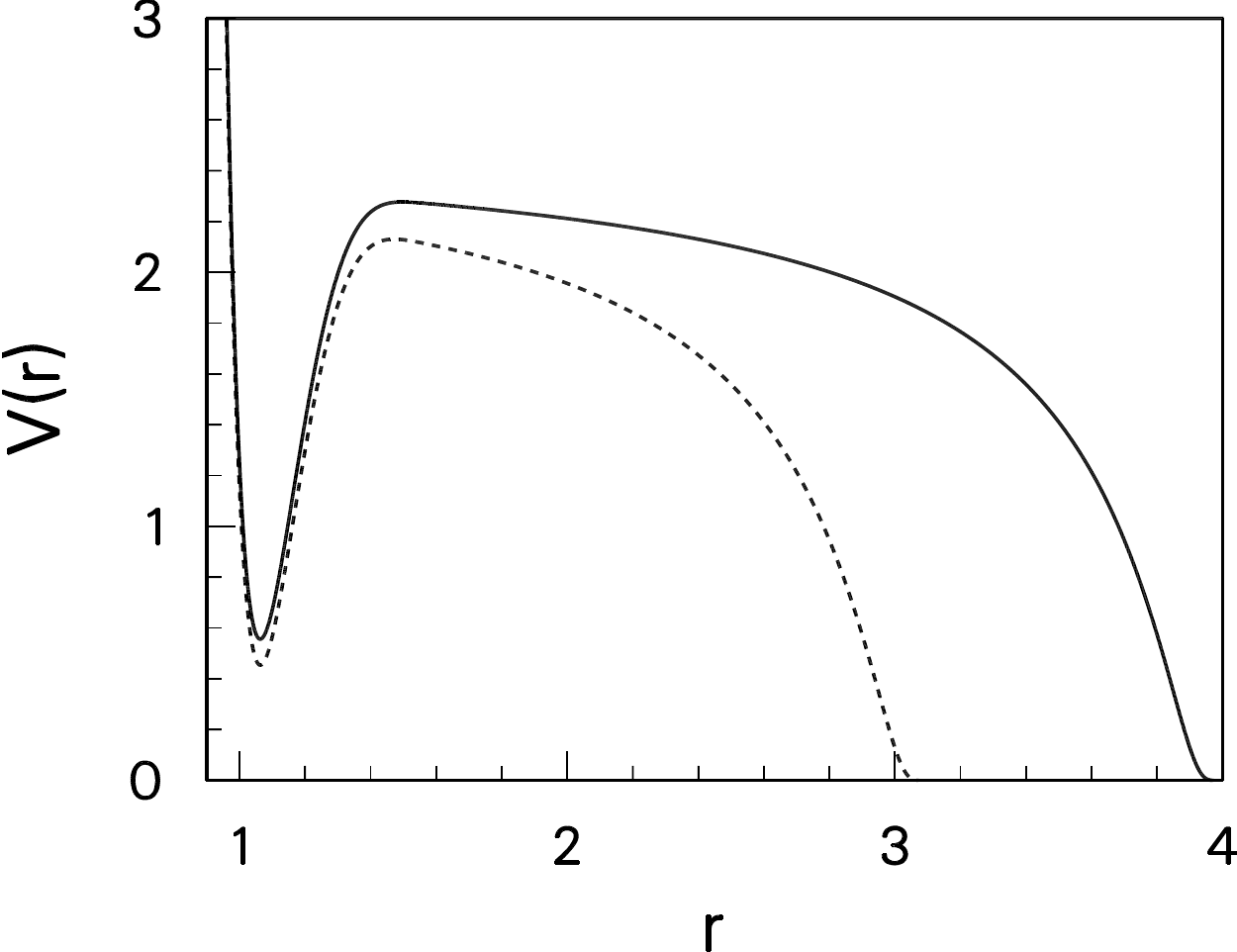}
\caption{ Pair potentials used in this study. Solid line, $V_1$;
  dashed line, $V_2$.}
\label{fig1}
\end{figure}

Each of the investigated molecular-dynamics models was composed of
16384 identical particles confined to a cubic box with periodic
boundary conditions. They used two pair potentials, $V_1$and $V_2$,
shown in Fig.\ref{fig1}.  The potential energy for two particles
separated by
the distance $r$ is described by the functional form:\\
\begin{eqnarray}  
V(r)  = a_1  ( r^{-m} - d) H(r,b_1,c_1) + a_2 H(r,b_2,c_2)\\ \nonumber
H(r,b,c) = \left\{ \begin{array}{ll}
  \exp\left( \frac{b}{r-c} \right) & r < c\\ 
0 & r \geq c 
\end{array} 
\right.       
\end{eqnarray}

\begin{table} [h]
\begin{tabular}{ccccccccc}
 \hline 
\hline 

 & m  & $a_1$ & $b_1$ & $c_1$ & $a_2$ & $b_2$ & $c_2$ & $d$ \\

\hline 
$V_1$ & 12 \space & 113 \space & 2.8 \space & 1.75 \space & 2.57 \space & 0.3 \space &
4.0 \space & 1.4\\
\hline 
\hline 
$V_2$ & 12 \space & 113 \space & 2.8 \space & 1.75 \space & 2.57 \space & 0.3 \space &
3.1 \space & 1.4\\
\hline 
\end{tabular}
\caption{Values of the  parameters for the pair potentials.} 
\label{table1}
\end{table}

The values of the parameters are presented in Table \ref{table1}.  The
reduced units used in this simulation are those used in the definition
of the potentials.
\begin{figure} [h!]  
\includegraphics[width=5.cm]{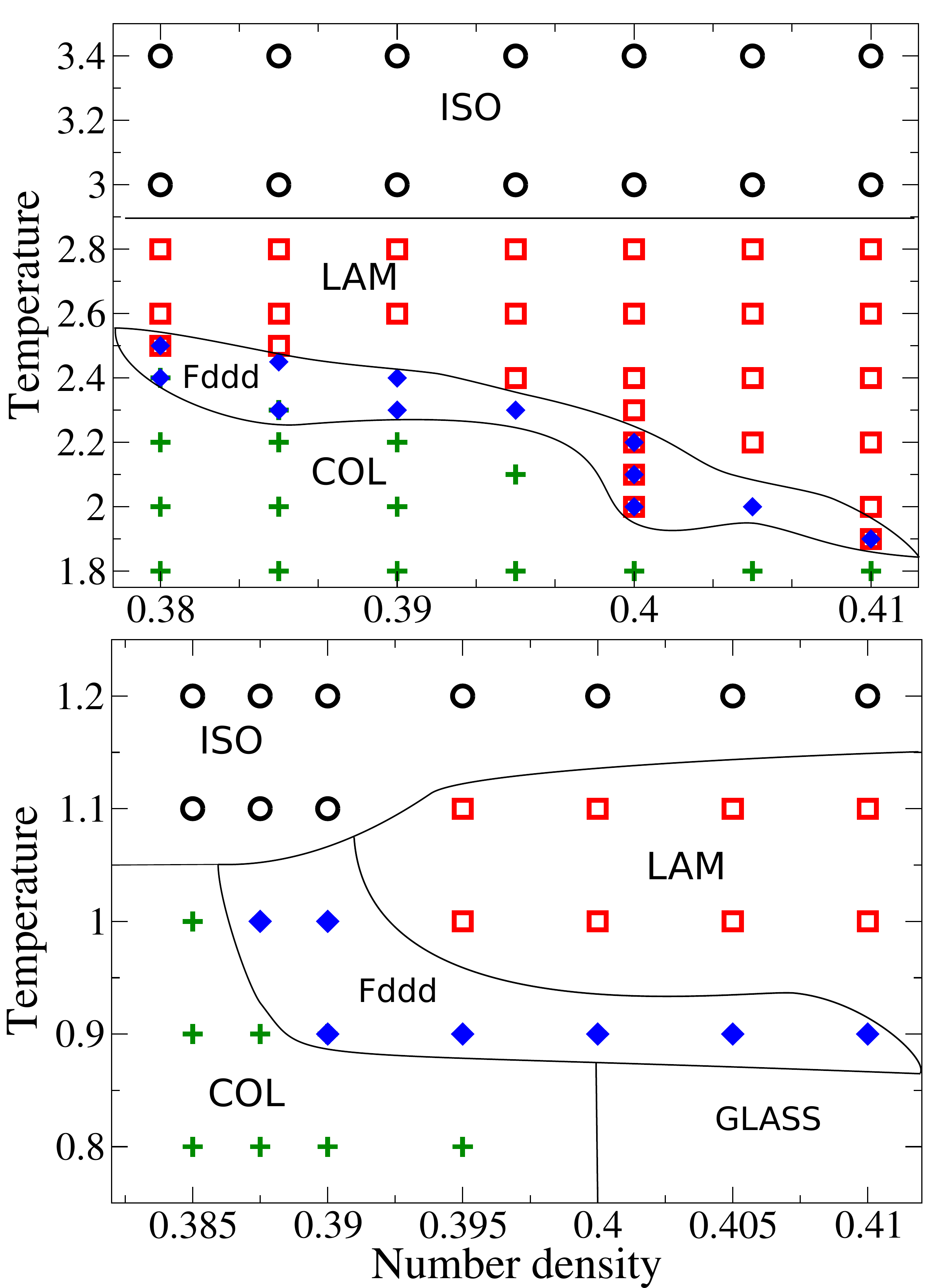}
\caption{ Phase diagrams of the symulated systems. Top and bottom,
  respectively, System I and System II. Open circles: isotropic liquid
(ISO). Open squares: lamellar phase (LAM). Diamomds: $Fddd$. Crosses:
columnar phase (COL).}
\label{fig2}
\end{figure}\\ 
This functional form of the pair potential is consistent with the
general form of the SALR potentials and it was earlier found to produce
columnar and lamellar mesophases \cite{col,hex}. In $V_1$ the
long-range repulsion is extended to a significantly larger distance
than in $V_2$. In the following the systems using potentials $V_1$ and
$V_2$ will be referred to as System I and System II, respectively.
\begin{figure}  [h!] 

\includegraphics[width=8.cm]{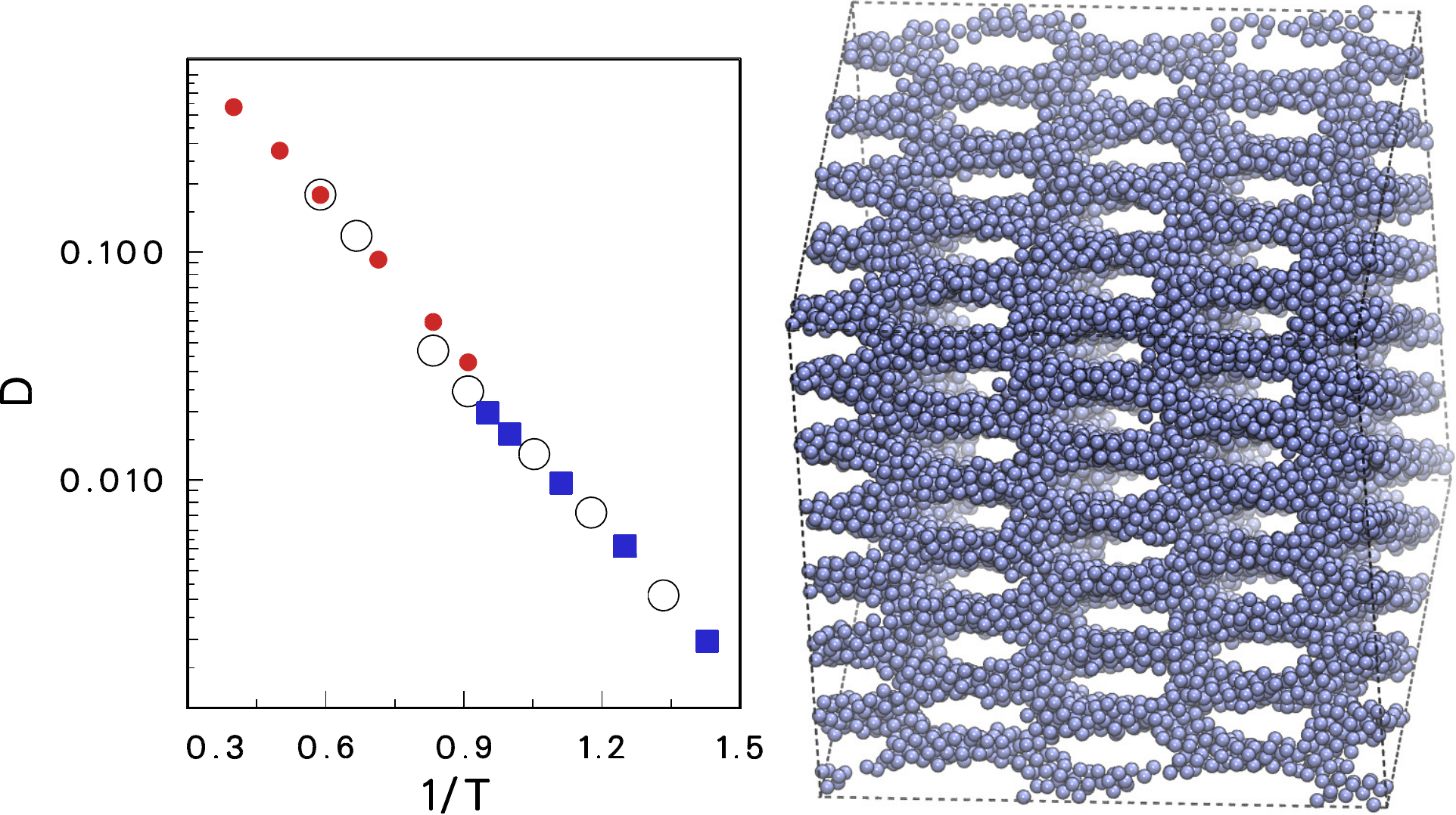}
\caption{ Left: Arrhenius plot for System II. Dots: isotropic liquid;
  squares: $Fddd$ phase, cooling; open circles: $Fddd$ phase,
  reheating. Right: an instantaneous particle configuration of the
  $Fddd$ phase formed in System I.}
\label{fig3}
\end{figure}

We explored the phase behaviour of both systems by performing
isochoric coolings within a range of densities starting from the
isotropic liquid state equilibrated at high temperature. The cooling
to the targeted points of the phase diagram was performed by
independent discontinuous quenching steps with comprehensive
equilibration following each step. Under cooling, both systems were
found to perform liquid-gas microphase separation transitions
producing equilibrium periodic morphologies which were identified by
visual inspection of the instantaneous particle configurations.

The density-temperature phase diagrams of the simulated systems are
shown in Fig.\ref{fig2}.  Besides the domains of classical
morphologies, lamellae and hexagonally packed cylinders, both systems
exhibit domains where the microphase separation transitions produce
equilibrium triply periodic networks with the $Fddd$ space group
symmetry. An instantaneous particle configuration of this microphase
produced by System I is shown in Fig. 5. In the following we present a
detailed analysis of the structure and the dynamical properties of the
simulated $Fddd$ morphologies.

For both systems the domains where the self-assembly of equilibrium
$Fddd$ microphase was observed extend within a significant range of
densities. We also observe a difference in the systems' phase
behaviour. For System I, the $Fddd$ domain is separated from the
uniform liquid domain by a domain of lamellar phase. Therefore, it was
only possible to produce the $Fddd$ self-assembly by discontinuously
quenching it from the equilibrium liquid state to a targeted range of
temperatures, whereas System II performed the liquid-$Fddd$ transition
under a continuous cooling.  That transition was also found to be
reversible when the system was reheated in a similar continuous
manner.  This phase behaviour demonstrates that the observed $Fddd$
microphases are thermodynamically robust equilibrium phases. We note
that there have been no phase transformations observed when either of
the two systems was heated or cooled across the boundaries separating
the domains of the ordered periodic microphases.  This can be
explained by the close degeneracy of the free energies of different
microphase morphologies at the same thermodynamic parameters
\cite{leb,edelman}. This degeneracy can also account for the significant
metastability area between the lamellar and $Fddd$ domains in the
phase diagram of System I where both phases have been observed
self-assembling.

\begin{figure} [h!] 
\includegraphics[width=8.1cm]{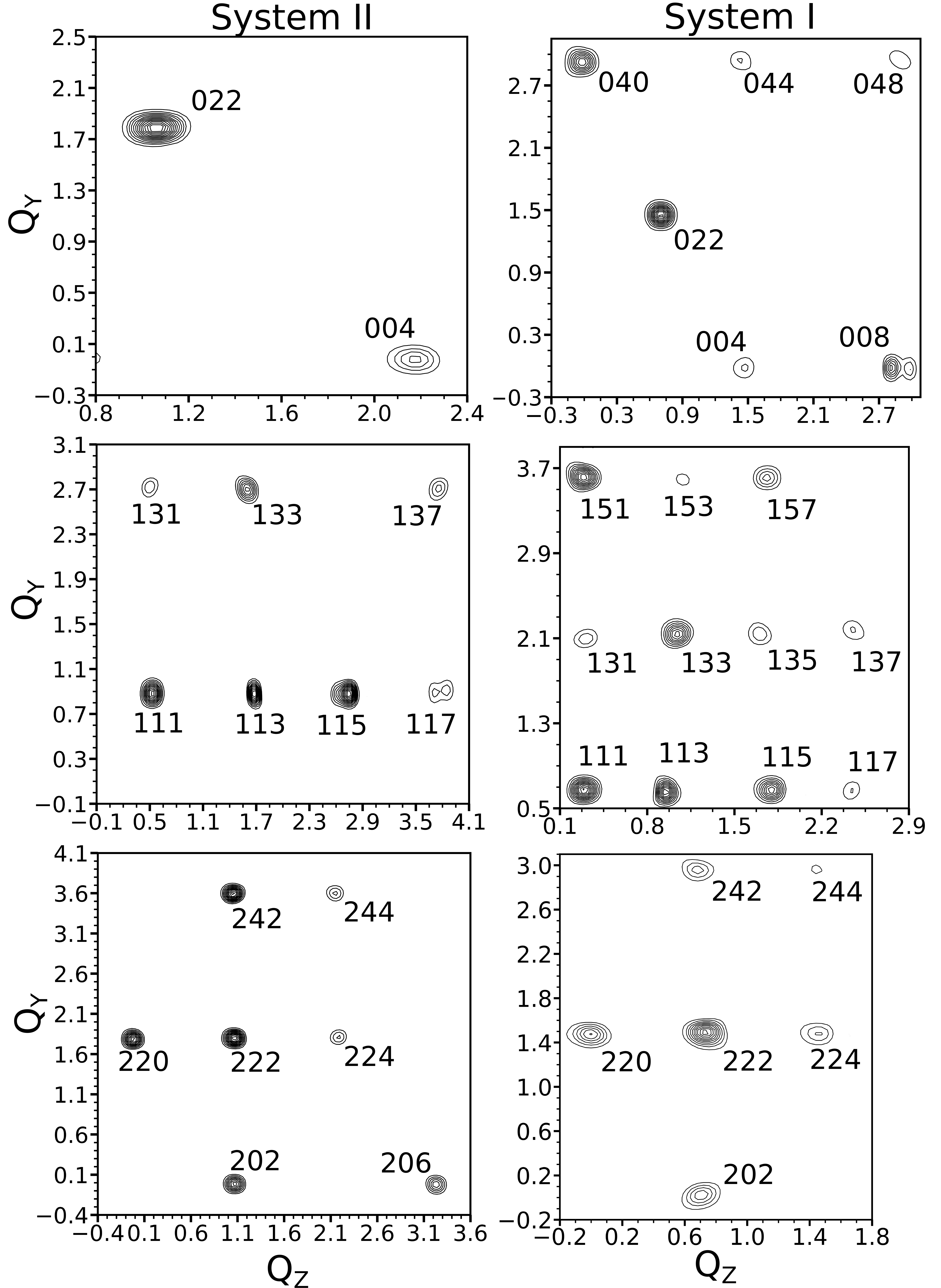}
\caption{Isointensity contour plots of the structure factor for the
  two systems.}
\label{fig4}
\end{figure}

\begin{figure}  
\center
\includegraphics[width=8.cm]{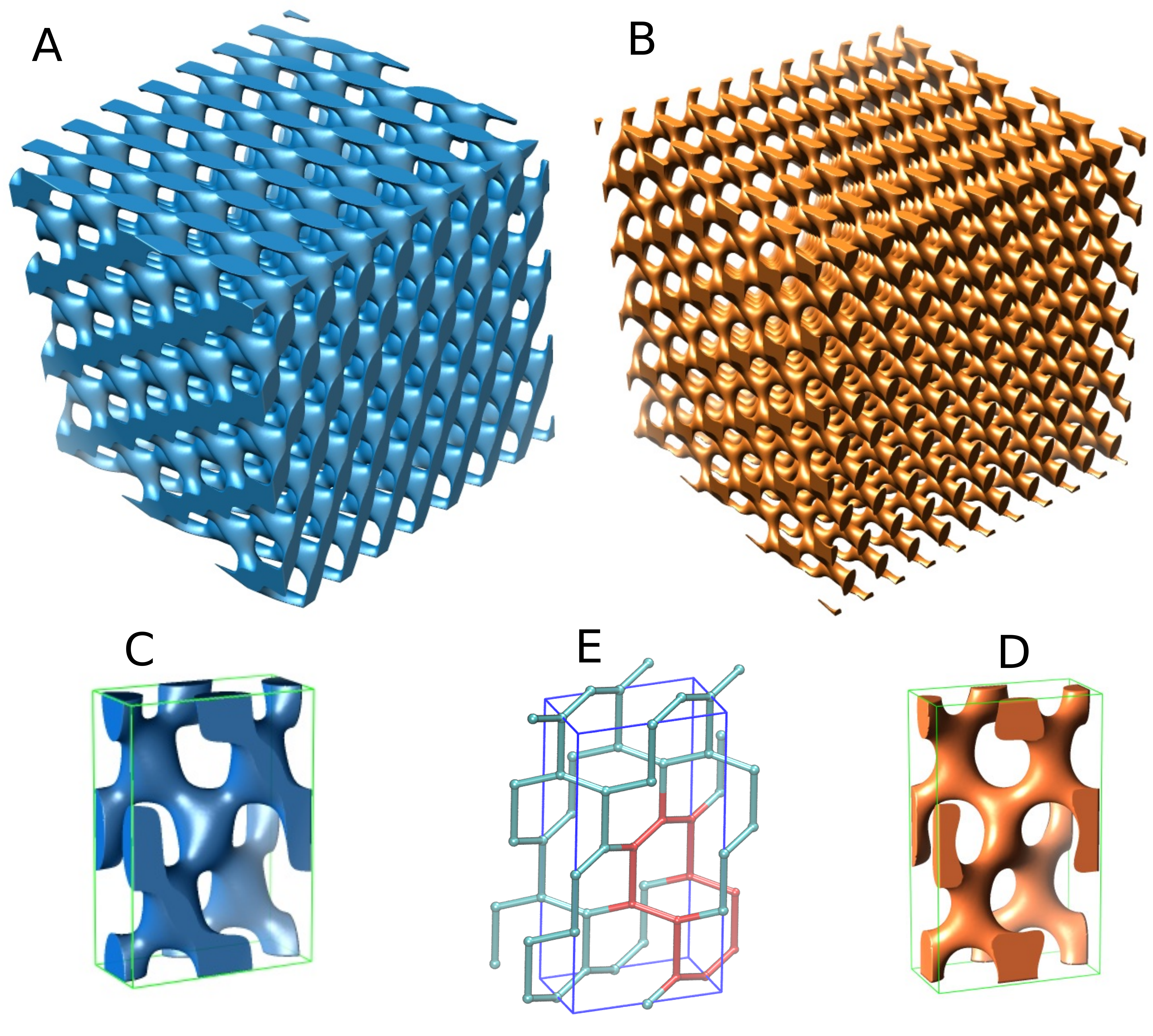}
\caption{A and B: the density isosurfaces for the systems I and II,
  respectively. C and D, respectively: the unit cells
  for the Systems I and II cut from the density isosurfaces. E: ball-
  and-stick model for System I, The (10,3) ring is highlighted.  }
\label{fig5}
\end{figure}
\twocolumngrid 

It has been argued that the diffusion of the constituent polymer
chains along the separating surface in the triply periodic networks
formed by block copolymers is blocked by the strong separation effects
\cite{meuler,epps}. By contrast, the particles in the simulated $Fddd$
morphologies demonstrate liquid-like diffusion.  Fig.\ref{fig3} shows
the Arrhenius plot of the diffusion rate in System II within the
relevant range of temperatures. No significant difference in either
the diffusion rate or the activation energy is observed upon the
transition between the uniform liquid and the $Fddd$ demonstrating the
fluid nature of the observed microphase, see also \cite{movie} We can
thus conclude that this transition is a microphase separation within
the liquid-gas spinodal domain as conjectured by Lebowitz and Penrose
\cite{leb}.

The structure factor of a system of $N$ particles is defined as:
\begin{equation} 
  S({\bf Q})= \frac{1}{N}  \left | \, \sum_{j=1}^{N} e^{i{\bf Q}{\bf r}_j} \right|^2\\
\end{equation}

where $\bf{r}_j$ are the particle positions.  To remove thermal
fluctuations, the instantaneous particle configurations considered
here have been subjected to the steepest-discent
minimisation. Fig.\ref{fig4} shows the maxima of $S({\bf Q})$ for both
systems in three characteristic $\b{Q}$-planes labelled by Miller
indices $hkl$. These patterns are consistent with the the small-angle
X-ray scattering measurements on the $Fddd$ networks found in block
copolymers \cite{meuler,jung}. We note that this is the first 3D
observed diffraction pattern of a $Fddd$ network.

The positions of the observed maxima of $S({\bf Q})$ are related to
the orthorombic unit cell parameters $a,b,c$ as ${\bf
  Q}_{hkl}=2\pi(h/a, k/b, l/c)$. For the system I we found $(a:b:c) =
(1:2.0:4.0)$, $a=4.3$. For the system II the respective results are
$(a:b:c) = (1:1.99:3.32)$, $a=3.49$. The parameters' ratios for the
System II reproduce those found in the $Fddd$ structures formed by
block copolymers \cite{epps, jung} within the margin of the accuracy
of measurements. For the System I, the ratio $c/a$ appreciably
exceeds the experimentaly observed value \cite{epps}, which can
possibly be ascribed to the impact of the boundary conditions.

The $Fddd$ unit-cell parameters are intimately related to the problem
of thermodynamic stability of this orthorhombic microphase relative to
the cubically symmetric gyroids. This symmetry breaking can be
understood in terms of weak segregation theory \cite{leibler}. The
Landau expansion of the free energy in terms of composition modes
demonstrates spinodal instability for the modes corresponding to the
lattice parameters obeying the relation: $(a:b:c)=(1:2:2\sqrt(3))$
\cite{tyler}. The lattice parameters for the system II agree with this
relation quite closely, whereas for the system I the ratio $c/a=2$
deviates from the one predicted by the theory.  This can possibly
account for the difference in their phase behaviour: System II
demonstrated direct thermoreversible transformation between the
isotropic liquid phase and the $Fddd$ phase whereas for System I these
two phases are separated by a domain of lamellar phase. Also, that
system's phase diagram demonstrated considerable overlap between the
LAM and the $Fddd$ domains. The deviation of the $c/a$ ratio in System
I from the value minimising the free energy can possibly be explained
by the constraints imposed by the periodic boundaries. On the other
hand, the fact that despite this distortion the system still
self-assemble into the $Fddd$ network can be viewed as the evidence of
the robust nature of this transition.

It is convenient to present bicontinuous periodic morphologies in
terms of the triply periodic surfaces segregating the two percolating
domains. For the present morphology formed as a result of the
liquid-gas microphase separation this would be the density
isosurface. For that purpose, a continuous perfectly periodic
real-space density distribution has been produced by the inverse
Fourier transformation of the calculated structure factor, with the
latter's width constrained by a Gaussian window function of an
appropriately chosen width. In this way sufficiently smooth uniform
density isosurfaces have beeen produced eliminating the effects of the
local fluctuations of the density distribution. These isosurfaces are
shown in Fig.\ref{fig5} which also shows the unit cells of the
simulated $Fddd$ periodic networks as cut fragments of the density
isosurfaces.

The topology of triply periodic networks is commonly analysed in terms
of skeletal graphs, ball-and-stick models \cite{jung}. Such a model
for the unit cell of System I is shown in Fig.\ref{fig5}E. The stick
lengths and positions have been produced as the best fits to the
isosurface.  According to the classification of Wells \cite{wells},
triply periodic nets can be viewed as the tilings of space by
closed loops of $n$ nodes with $p$ connections each, denoted as
$(n,p)$. The $Fddd$ nets were concluded to be composed of $(10,3)$
loops \cite{meuler}. Such a loop is highlighted in the ball-and-stick
model shown in Fig.\ref{fig5}E. Like in the experimentally observed
$Fddd$ networks \cite{jung}, the three-fold symmetry of the nodes is
found to be broken with longer sticks oriented along the $c$ axis.

We note that the distance between the longer sticks coincides with the
parameter $a$ of the unit cell. For both systems this distance is
consistent with the long repulsion range of the respective
potential. Thus we can rationalise the observed self-assembly of the
$Fddd$ network in a simple system of particles as follows: the
attraction part of the pair potential favours condensation whereas the
range of its long repulsion part defines the basic periodicity of the
network, its symmetry arises from the Landau free energy
minimum as a result of the superposition of composition modes
\cite{leibler, tyler}

We conclude with the following remarks.\\
1. It has been postulated that the entropic effects of polymer chain
stretching are primarily responsible for the network morphologies in
block copolymers. In particular they were suggested to account for the
3-fold connectors universally observed in these networks. The finding
of this topology in simple systems suggests that
a more general mechanism is responsible for its formation.\\
2. This is the first particle simulation of the $Fddd$ network. The
detailed particle-level information about the geometry of its triply
periodic domains can be expected to advance the development of the
$Fddd$ minimal surface which hasn't
been reported so far.\\
3. The self-assembly of the $Fddd$ network in two distinctly different
simple systems within a range of densities evidences the robust nature
of this phase transition.  This form of the pair potential can be
expected to produce other triply periodic networks in simple systems
including colloids where the interparticle interactions can be tuned
to approximate it.

We are grateful to Prof. Terasaki for useful discussions.
These simulations used GROMACS software.

\end{document}